\documentclass[prd,showpacs,twocolumn,preprintnumbers]{revtex4-1}
\usepackage{mathptmx}
\usepackage[english]{babel}
\usepackage{amssymb}
\usepackage{amsfonts}
\usepackage{amsmath}
\usepackage{eucal}
\usepackage{graphicx}
\usepackage{epsfig}
\usepackage{slashed}
\newcommand{\be}{\begin{equation}} \newcommand{\ee}{\end{equation}}
\newcommand{\bea}{\begin{eqnarray}} \newcommand{\eea}{\end{eqnarray}}
\newcommand{\beann}{\begin{eqnarray*}}  \newcommand{\eeann}{\end{eqnarray*}}
\newcommand{\bfig}{\begin{figure}} \newcommand{\efig}{\end{figure}}
\newcommand{\ba}{\begin{array}} \newcommand{\ea}{\end{array}}
\newcommand{\bcen}{\begin{center}} \newcommand{\ecen}{\end{center}}
\newcommand{\btab}{\begin{tabular}} \newcommand{\etab}{\end{tabular}}

\newtheorem{Proposition}{Proposition}[section]

\newtheorem{Theorem}{Theorem}[section]
\newtheorem{Lemma}{Lemma}[section]
\newtheorem{Corrolary}{Corrolary}[section]

\newcommand{\bp}{\begin{Proposition}}	\newcommand{\ep}{\end{Proposition}}
\newcommand{\bt}{\begin{Theorem}}	\newcommand{\et}{\end{Theorem}}
\newcommand{\bl}{\begin{Lemma}}		\newcommand{\el}{\end{Lemma}}
\newcommand{\bc}{\begin{Corrolary}}	\newcommand{\ec}{\end{Corrolary}}
\begin{document}

%%%%%%%%%%%%%%%%%%%%%%%%%%%%%%%%%%%%%%
%%%%%%%%%%%%% TITLEPAGE %%%%%%%%%%%%%%
%%%%%%%%%%%%%%%%%%%%%%%%%%%%%%%%%%%%%%
\title{A simple holographic scenario for gapped quenches}

\author{Esperanza Lopez and Guillermo Milans del Bosch}%\email{esperanza.lopez@uam.es}
\affiliation{Instituto de F\'{\i}sica Te\'orica UAM/CSIC, C/ Nicol\'as Cabrera
13-15,\\
Universidad Aut\'onoma de Madrid, Cantoblanco, 28049 Madrid, Spain}

\begin{abstract}
We construct gravitational backgrounds dual to a family of field theories parameterized by a relevant coupling. They combine a non-trivial scalar field profile with a naked singularity. The naked singularity is necessary to
preserve Lorentz invariance along the boundary directions. The singularity is however excised by introducing an infrared cutoff in the geometry. The holographic dictionary associated to 
the infrared boundary is developed. We implement quenches between two different values of the coupling. This requires considering time dependent boundary conditions for the scalar field both at the AdS boundary and the infrared wall.
\end{abstract}

\preprint{IFT-UAM/CSIC-17-002}
\maketitle
%
%%%%%%%%%%%%%%%%%%%%%%%%%%%%%%%%%%%%%%
%%%%%%%%%%%% INTRODUCTION %%%%%%%%%%%%
%%%%%%%%%%%%%%%%%%%%%%%%%%%%%%%%%%%%%%

{\it Introduction.} 
Modeling quantum quenches in a holographic setup has attracted considerable attention in the last years. A remarkable success has been achieved in reproducing important aspects of the universal dynamics of quenches \cite{AbajoArrastia:2010yt}-\cite{Liu:2013iza}. However most models lack some of the defining characteristics of quenches. Notably, they simulate an injection of energy in the system without real change in the hamiltonian.

In this note we want to present a simple holographic model of a quench modifying the infrared physics. 
With this aim, we search for the gravitational dual to a family of d-dimensional QFT's parameterized by a relevant coupling. As a main input, the ground state for any value of the coupling is required to be Lorentz invariant. 
We pursue the minimal scenario, involving Einstein gravity coupled to a real scalar field. The possibility of extra compactified dimensions is excluded. 

We use the following ansatz for the ground state metrics 
\be
ds^2=  \frac{1}{z^2}\left( {dz^2 \over A(z)} - dt^2 +  d {\vec x}_{d-1}^{\, 2}\right) \, .
\label{metricgr}
\ee
Setting $8 \pi G\!=\!d\!-\!1$, the equations of motion are
\be
{z \over 2} \,A'= d (A-1) + 2 V(\phi)\, , \hspace{.5cm}    A'=2 \, z A {\phi'}^2 \, .
\label{eomd}
\ee 
These equations have two integration constants, which can be related to the coefficients of the two independent scalar modes. Asking for regularity of the geometry links their values, allowing to interpret one of them as a QFT coupling and the other as the expectation value of the sourced operator. 
Regular solutions of \eqref{eomd}, whose existence depends on the scalar potential, describe RG flows into an infrared fixed point independent of the integration constants. 
All other solutions run into naked singularities. Considering naked singularities raises a number of serious problems. 
There is no condition that relates the two integration constants, challenging the usual holographic dictionary. 
Related to this, Lorentz invariant metrics are not minimal energy solutions when only one integration constant is fixed at the AdS boundary. Actually there are solutions of arbitrary negative energy. 

These issues admit a simple, albeit crude solution, by introducing an infrared cutoff in the geometry. This creates a new boundary and renders  natural to interpret both integration constants from \eqref{eomd} as couplings.
Fixing the two couplings solves the vacuum stability problem \cite{Horowitz:1995ta}.
Moreover regions of high curvature, 	which bring outside the regime of validity of classical gravity, are excised. 

AdS with an infrared hard wall is a well known rough holographic model for confining theories \cite{Erlich:2005qh}. The new ingredient in this paper is to consider the hard wall as a regularizing element, while the infrared physics will be linked to the strength of the naked singularity.

\vspace{3mm}

{\it Static backgrounds.} 
We will explore the proposed scenario with a vanishing scalar potential. 
Equations \eqref{eomd} can then be solved analytically, with the result
\bea
A(z) &=& 1+ \alpha^2 z^{2d} \; , \label{A} \\
\phi(z) &=& \beta+ d^{-1/2}\; {\rm arcsinh} (\alpha z^d)  \, , \label{phi}
\eea
where $\alpha$ and $\beta$ are arbitrary constants. $\beta$ represents a global shift in the value of the scalar, which is of no physical consequence when $V(\phi)\!=\!0$. $\alpha$ induces a non-trivial scalar profile
\be
\alpha=z_0^{-d} \sinh ( \sqrt{d} \, \Delta \phi ) \, .
\ee
with $z_0$ denoting the radial position of the wall, and $\Delta\phi\!=\!\phi_0\!-\!\phi_\infty$ the variation of the scalar field between the wall and the AdS boundary. If extended beyond the infrared cutoff, all backgrounds with $\Delta \phi\!\neq\!0$ have a naked singularity. 

\begin{figure}[h]
\begin{center}
\includegraphics[width=4cm]{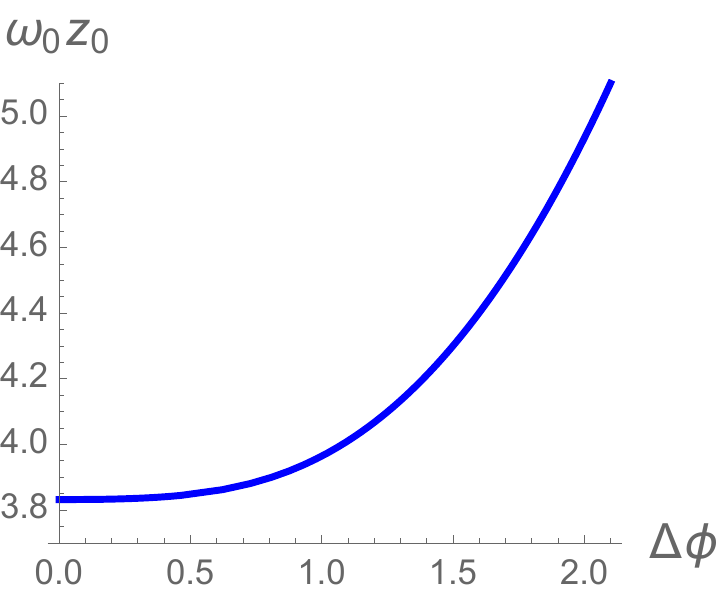}~~~~~ 
\includegraphics[width=4cm]{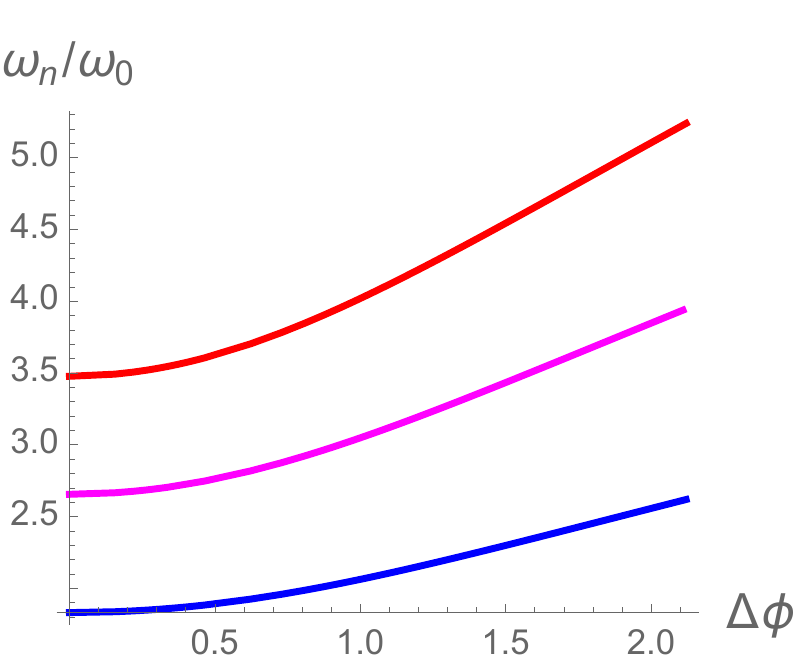}~~~~~
\end{center}
\vspace{-2mm}
\caption{\label{fig:gap} Left: Frequency of the fundamental scalar mode. Right: Ratio of the next three normal frequencies to the fundamental one.}
\end{figure}
Holography interprets the harmonic modes of bulk fields as excitations in the dual QFT. In Fig.\ref{fig:gap} we have plotted the frequency of the lower scalar modes along the family \eqref{A}-\eqref{phi} for $d\!=\!2$. 
When the radial variation of scalar profile is small, the spectrum is determined by $z_0$ . 
The spectrum becomes instead ruled by $\Delta \phi$ for larger values of this parameter. Fig.\ref{fig:gap}a shows that the mass gap, holographically given by $\omega_0$, grows with $\Delta \phi$ and implies that this is a relevant coupling.
Interestingly the ratio of higher normal frequencies to the fundamental one shows an approximate linear growth, see Fig.\ref{fig:gap}b. Hence the infrared physics associated to the family \eqref{A}-\eqref{phi} does not differ by a mere rescaling.
The lowest excitation becomes increasingly separated from the rest the larger is $\Delta \phi$.

\vspace{1mm}

{\it Modelling a quantum quench.} 
We want to model a global quench between QFT's whose ground states are in the family \eqref{A}-\eqref{phi}. 
A convenient ansatz for the associated metric is
\be
ds^2 = \frac{1}{z^2}\left( - A(t,z) e^{-2\delta(t,z)} dt^2 + {dz^2 \over A(t,z)} +  d {\vec x}^{\,2}_{d-1}\right) \, .
\label{metric}
\ee
For vanishing scalar potential, the equations of motion are 
\bea
\dot\Phi &=& \left( A e^{-\delta} \Pi\right)' ~,~~~~~\dot \Pi = z^{d-1} \!\! \left( z^{1-d} A e^{-\delta} \Phi \right)' \, ,\label{eqforphi} \\
\delta' &=& z  \,(\Phi^2\! + \Pi^2) ~,~~~~  A'=z  \, A\,  (\Phi^2 \!+ \Pi^2)+ {d \over z} (A-1) \, , \;\;\; \label{eqforA} 
\eea
with $\Phi\!=\!\phi'$ and $\Pi\!=\!A^{-1} e^\delta \dot{\phi}$ encoding the radial and time scalar derivatives.
Solving the equations of motion requires giving a set of initial data together with boundary data at asymptotic AdS and the infrared wall. As boundary data, we will allow for time dependent profiles $\phi_\infty(t)\!=\!\phi(t,0)$ and $\phi_0(t)\!=\!\phi(t,z_0)$. Dynamical processes triggered by $\phi_\infty(t)$ in the hard wall setup were studied in \cite{Craps:2013iaa}. The possibility of imposing a
time dependent scalar profile at the wall has been considered in \cite{daSilva:2016nah}.
 
The family \eqref{A}-\eqref{phi} does not exhaust the set of static solutions to our gravity system. In general they break Lorentz invariance and are described by the ansatz \eqref{metric}. Their energy density can be read from the asymptotic expansion $A\!=\!1\!-\!2M z^d\!+\!\dots$. It is then clear that all solutions \eqref{A}-\eqref{phi} have zero mass. For static solutions
\be
M={1\over 2}z_0^{-d}  (1-A_0) + {1 \over 2}\! \int_0^{z_0} z^{1-d} A \Phi^2 dz  \, ,
\label{mass}
\ee
with $A_0\!=\!A(z_0)$.
The first term represents the contribution to the total energy from the geometry hidden by the infrared cutoff. When this part encloses a naked singularity it can be arbitrarily negative. On the contrary, the second term is always positive for solutions without horizons. Therefore naked singularities are a crucial ingredient for obtaining Lorentz invariant backgrounds with non-trivial scalar profiles.

\begin{figure}[h]
\begin{center}
\includegraphics[width=4.2cm]{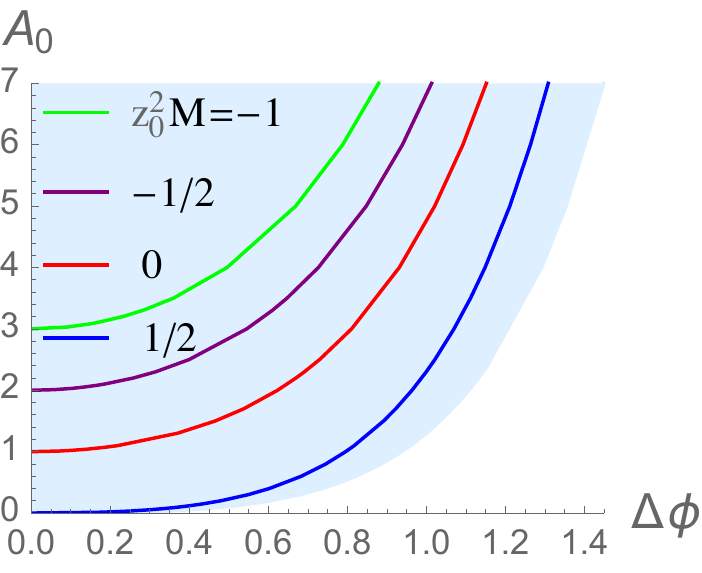}~~~
\includegraphics[width=4.2cm]{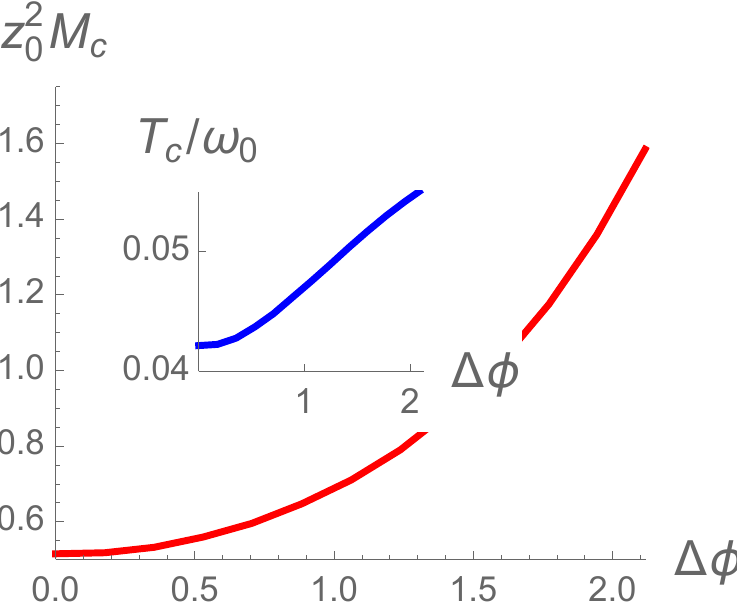}
\vspace{-5mm}
\end{center}
\caption{\label{fig:phase} Left: Shadowed in blue, couplings admitting static solutions without horizons for $d\!=\!2$. Equal mass curves are highlighted. Right: Threshold mass for collapse of pulses \eqref{mcol}. Inset: Ratio of the temperature of the black hole at threshold to the mass gap.}
\end{figure}

Up to a trivial global shift in the scalar, static solutions are parameterized by $\Delta \phi$ and $A_0$. 
The backgrounds \eqref{A}-\eqref{phi} define the codimension one subset 
\be
A_0=\cosh^2 (\sqrt{d} \, \Delta \phi ) \, ,
\label{mzero}
\ee
see red line in Fig.\ref{fig:phase}.
Since $A_0$ is a boundary data, it is natural to also interpret it as a QFT coupling. The unique static solution without horizons for $\Delta \phi$ and $A_0$ in the shaded region of Fig.\ref{fig:phase}a, represents the ground state of the associated QFT \cite{daSilva:2016nah}.

We consider that the QFT before the quench is in the ground state for chosen couplings in the subset \eqref{mzero}. Acting on the boundary values such that $\phi_\infty$ changes while $\phi_0$ remains constant,
clearly brings outside \eqref{mzero}. The same actually happens  in the opposite case. When $\phi_\infty$ is kept constant, the equations of motion ensure the conservation of total mass.  A time dependent $\phi_0$ generates a scalar pulse that enters the geometry at the infrared boundary. Unless the time variation is adiabatic, this pulse induces an excited state in the final QFT. The value of $A_0$ will then adjust such that the total energy is conserved. Namely, if the initial theory belongs to the Lorentz invariant subset, the final one will have negative ground state energy. In order to model a quench between theories in \eqref{mzero}, the wall profile $\phi_0(t)$ needs to be combined with an energy injection into the system. This can only happen at the AdS boundary, induced by a non-trivial $\phi_\infty(t)$.

The initial state will be taken to have vanishing $\phi_\infty$, $\phi_0\!=\!{\bar \phi}$ and  $A_0$ satisfying \eqref{mzero}.  We shall choose the wall profile
\be
\phi_0(t)={\bar \phi} +{1 \over 2} \eta \Big(1+\tanh{t \over a} \Big) \, .
\label{quench}
\ee
This models a quench with a finite time span controlled by the parameter $a$. After the quench $\phi_0\!=\!{\bar \phi}\!+\!\eta$, while $A_0$ will be fixed by the conservation of energy at the wall.
Any $\phi_\infty(t)$ which fulfills \eqref{mzero} at late times, ensures that the final QFT will have Lorentz invariant couplings. 
We do not want however that the quench follows an arbitrary path in the coupling space of Fig.\ref{fig:phase}a. We aim to only act on the combined coupling that moves along the $M\!=\!0$ line. This involves a tuned variation of the scalar field at the wall and the AdS boundary, which the absence of time-like Killing vector renders unclear how to implement.
In the following we will assume that the diagonal time coordinate in \eqref{metric} provides a reasonable way to project the value of $\phi_0$ onto the dual QFT. Hence we require \eqref{mzero} to hold at each constant time slice. 

\vspace{2mm}

{\it Numerical results.} The central characteristic of the dynamics generated by \eqref{mzero}-\eqref{quench}, is whether or not it will generate a horizon. In the affirmative case, the end point of the evolution is a Schwarzchild black hole trapping the total mass. This represents a unitary process in the dual QFT leading to thermalization  \cite{AbajoArrastia:2010yt}\cite{Takayanagi:2010wp}. Those that do not form a horizon result in a scalar pulse that bounces forever between AdS boundary and wall \cite{Craps:2013iaa}. 
Bouncing geometries provide the holographic counterpart to periodic reconstructions of quantum correlations in the dual field theory \cite{Abajo-Arrastia:2014fma}, known as quantum revivals \cite{Robinett2004}.

The only topological obstruction to the formation of a horizon in our setup is the presence of the wall, enforcing $M z_0^2\!>\!1/2$. A first question then is whether the typical scale triggering fast thermalization is set by $z_0$ or depends on the $\Delta \phi$.
Before studying quenches, we analyze the infall of a scalar shell modelling an energy injection without variation of the hamiltonian. We restrict in the following to $d\!=\!2$ for numerics.
Since the shape of the pulse influences the evolution, we consider a typical shell, radially localized and of gaussian form 
\be
\Pi(t\!=\!0) \propto z^2 e^{-{1 \over \sigma^2} \tan^2 ({\pi z \over 2 z_0}) } \; , 
\label{mcol}
\ee
with $\sigma\!=\!0.1$ and  $ \Phi(t\!=\!0)$ in the family \eqref{A}-\eqref{phi}.
The threshold mass for gravitational collapse without bounces is plotted in Fig.\ref{fig:phase}b. It strongly grows with $\Delta \phi$, confirming the secondary role of the infrared wall. Using the hard wall as an auxiliary element, we are actually obtaining a basic model of a soft wall.

\begin{figure}[h]
\begin{center}
\includegraphics[width=4.2cm]{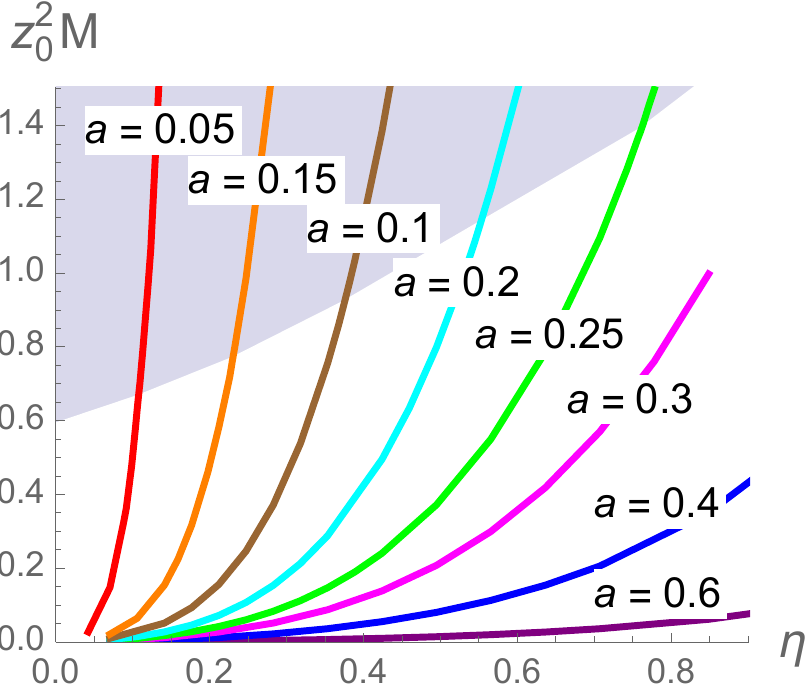}~
\includegraphics[width=4.6cm]{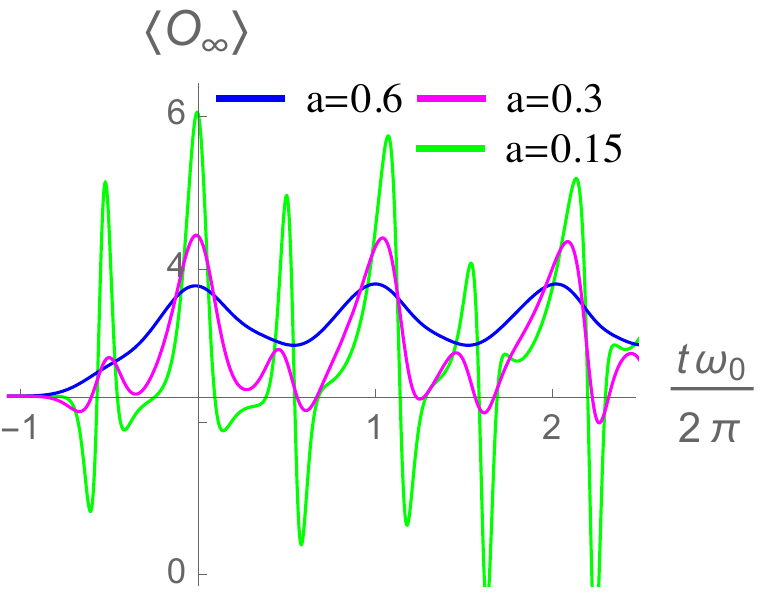}
\end{center}
\vspace{-2mm}
\caption{\label{fig:quench} Left: Energy density generated by \eqref{mzero}-\eqref{quench} for ${\bar \phi}\!=\!0.7$ and $d\!=\!2$. Right: $\langle {\cal O}_\infty \rangle$ for three quenches with different time spans.}
\end{figure}

We explore now the evolutions after a quench modelled by \eqref{mzero}-\eqref{quench} in $d\!=\!2$.
The quench will be applied to the Lorentz invariant background $\Delta \phi\!=\!{\bar \phi}\!=\!0.7$. At this value the infrared physics starts to be dominated by the hidden singularity instead of the wall position, see Fig.\ref{fig:gap}. 
We focus on $\eta\!>\!0$, and hence the quench will increase the mass gap. Fig.\ref{fig:quench}a shows the final energy density as a function $\eta$ for several values of the time span $a$. Its growth with $\eta$ is more pronounced the smaller is $a$. We have shaded in blue the parameters that lead to black hole formation. Processes where a horizon is generated after some bouncing cycles occupy just a small window on the boundary of the blue region. Otherwise we obtain geometries that keep bouncing as far as our simulation could go. 
Only sufficiently fast quenches, those with $a\!<\!0.25$ in the example of Fig.\ref{fig:quench}a, can generate enough energy density to trigger thermalization.

Bouncing geometries can be roughly divided in two types: standing and traveling waves. Standing waves project mainly on the fundamental harmonic of the static background associated to the final couplings. It is convenient to restore the natural mass units, $M \rightarrow {d-1 \over 8 \pi G} M$, with $G$ extremely small. According to the holographic dictionary, $1/G$ is proportional to the number of elementary degrees of freedom in the dual QFT. Hence $M$ translates into an energy density per species in field theory terms. Although the mass of  standing waves is much smaller than that required for collapse, it can be parametrically larger than $G$. 
Indeed quenches in Fig.\ref{fig:quench}a generate standing waves when $a\!\geq\!0.6$, having masses up to $M z_0^2 \approx 0.1$. 

Standing waves oscillate with the frequency of the mass gap, $\omega_0$. 
It is then natural to holographically identify them with coherent states of ${\vec k}\!=\!0$ modes of the lowest QFT excitation. 
Revivals with the same interpretation appear for example in the massive Schwinger model after a quench \cite{Buyens:2013yza}. The important difference in our case is their energy density. It can be much larger than the mass gap, proper in holographic models of a confining phase, ranging up to $O(1/G)$, close to the typical values in the plasma phase.
In spite of that, the physics driving thermalization does not refer to $\omega_0$.
This is illustrated in the inset of Fig.\ref{fig:phase}b. The temperature of the black hole at the collapse threshold for the gaussian pulses \eqref{mcol}, is well below the mass gap.

Traveling pulses exhibit radial localization and displacement. They represent in general partial revivals. They have larger masses, and the associated QFT states are thus expected to contain higher energy excitations and non-zero momentum modes. The former should be connected with the projection of narrow pulses on higher harmonic modes. 
The radial infall of a narrow shell has been related to the evolution of the separation between entangled excitations after a quench \cite{AbajoArrastia:2010yt}\cite{Balasubramanian:2010ce}\cite{Liu:2013iza}, the so-called horizon effect \cite{Calabrese:2005in}.
In this sense, radial displacement indicates the presence of non-zero momentum modes in the dual field theory state. Since the quench we are modelling is global, finite momentum modes can only be created in pairs. Fig.\ref{fig:gap}b shows that $2 \omega_0\!\approx\!\omega_1$ for a large range of couplings, 
explaining why radial localization and displacement appear at similar energies.

Contrary to standing waves, traveling configurations generated by \eqref{mzero}-\eqref{quench} are composed of two distinct sub-pulses, one entering from the AdS boundary and the other from the wall. This is clearly appreciated in the one-point functions. Fig.\ref{fig:quench}b shows the vev of the operator sourced by $\phi_\infty$ for three examples from Fig.\ref{fig:quench}a. We use a rescaled time such that the fundamental frequency for the final couplings is $2 \pi$. The oscillations of $\langle {\cal O}_\infty \rangle$ are plotted in blue for a slow quench, with $a\!=\!0.6$, resulting in a standing wave. A traveling configurations with two sub-pulses producing signals of similar magnitude is obtained for $a\!=\!0.15$ and shown in green. The effect of both sub-pulses superposes, giving rise to oscillations with roughly twice the fundamental frequency. In magenta we have an intermedium configuration, with a small boundary component. It is worth mentioning a slight increase in the period of oscillations between the $a\!=\!0.6$ and $a\!=\!0.15$ pulses. This is due to their different final energies: $M\!=\!0.002$ and $M\!=\!0.02$ respectively. The increase of the period with the energy is generic in holographic quenches, finding some analogues in the condensed matter literature \cite{Abajo-Arrastia:2014fma}.

\begin{figure}[h]
\begin{center}
\includegraphics[width=4.3cm]{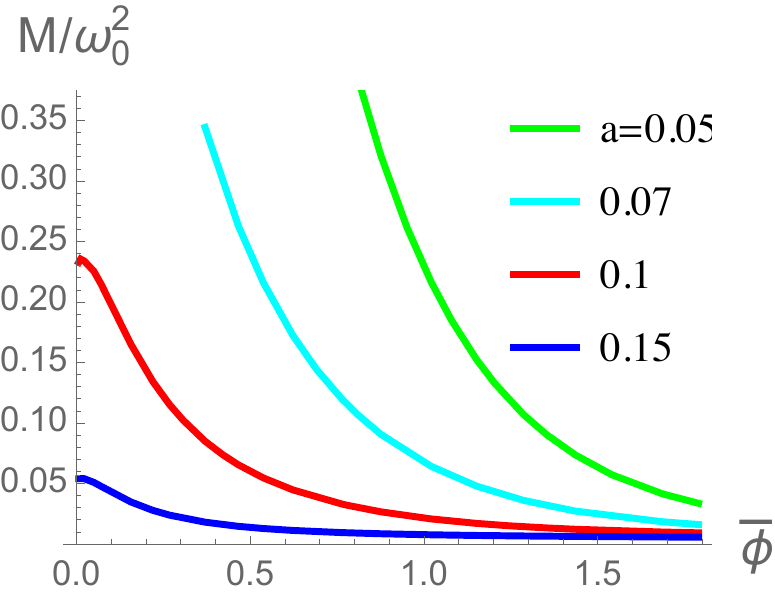}~~~
\includegraphics[width=4.3cm]{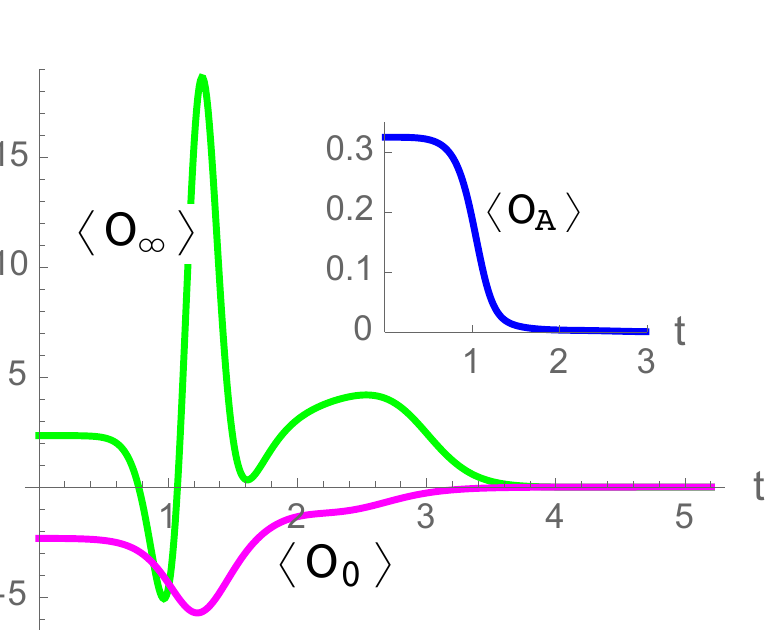}
\end{center}
\vspace{-2mm}
\caption{\label{fig:dev} Left: Energy density normalized by the square of the final mass gap in quenches with different initial coupling and $\eta\!=\!0.2$. Right: Evolution of one-point functions after a quench with $a\!=\!0.3$ and $\eta\!=\!1$, leading to thermalization.}
\vspace{-4mm}
\end{figure}

The distinction between fast and slow quenches should refer to the characteristic scale of the infrared physics. Slow quenches can be unambiguously defined as those producing standing or quasi-standing waves. We consider now quenches with fixed amplitude $\eta$ and time span $a$, but different initial coupling $\bar \phi$. Fig.\ref{fig:gap}a shows that the mass gap grows with the coupling. Therefore the quench should result in a collapsing shell, a bouncing pulse or a standing wave as we choose larger values of $\bar \phi$. Alternatively, the energy density in units of the final mass gap must be a monotonically decreasing function of $\bar \phi$. This quantity is plotted in Fig.\ref{fig:dev}a for $\eta\!=\!0.2$ and several small values of $a$, confirming the expected behavior. 

\vspace{2mm}

{\it One point functions at the wall.} We have assumed that the boundary values $\phi_0$ and $A_0$ relate to couplings with a well defined, local projection on the field theory time coordinate. The same as $\phi_\infty$, they should source local operators. We aim to determine their expectation values.

Symmetry under global shifts of the scalar field implies that only the difference $\Delta\phi\!=\!\phi_0\!-\!\phi_\infty$ is physically relevant. Hence the ground state expectation values of the operators ${\cal O}_0$ and ${\cal O}_\infty$ can not be independent. While the latter is dictated by the asymptotic expansions at the AdS boundary, the former has to depend on quantities evaluated at the  wall.
The scalar equation for static solutions reduce to $(z^{1-d} A e^{-\delta} \Phi)'\!=\!0$, implying
\be
\lim_{z\rightarrow 0}( z^{1-d} \Phi)  =z_0^{1-d} A_0 e^{-\delta_0} \Phi_0 \, ,
\label{vev}
\ee
where we have gauge fixed $t$ to be the proper time at the AdS boundary, {\it i.e.} $\delta_\infty\!=\!0$. 
The {\it lhs} is precisely  $\langle {\cal O}_\infty \rangle$ \cite{deHaro:2000vlm}. Defining $\langle {\cal O}_0 \rangle$ as minus the {\it rhs}, we obtain a relation of the desired form
\be
\langle {\cal O}_\infty \rangle\!+\! \langle {\cal O}_0 \rangle =0  \, .
\ee
The sign has been chosen such that the operator sourced by $\phi_\infty\!+\!\phi_0$ has a vanishing vev in the ground state. Notice that \eqref{vev} would not hold with $V(\phi)\!\neq\!0$, when neither a global shift on the scalar is a symmetry of the system.

The metric function $A$ satisfies the evolution equation
\be
{\dot A} =   2 z A  \Phi {\dot \phi} \, .
\label{mc}
\ee
The $z^d$ coefficient in the asymptotic expansion of $A$ determines the dual QFT energy density \cite{deHaro:2000vlm}. The previous equation implies ${\dot M}\!+{\dot \phi}_\infty \langle {\cal O}_\infty \rangle\!=\!0$. However the field theory Ward identities dictate a sum over all couplings, ${\dot M} \!+\!\sum {\dot \lambda}_i \langle {\cal O}_i \rangle\!=\!0$ \cite{deHaro:2000vlm}. It is then necessary that the contributions from $\phi_0$ and $A_0$ exactly cancel, which is the requirement of energy conservation at the wall. Using the above proposed value for $\langle {\cal O}_0 \rangle$,  \eqref{mc} at the wall can be rewritten as 
\be
{\dot \phi}_0 \langle {\cal O}_0 \rangle +  {1\over 2} {\dot A}_0  \, z_0^{-d}   e^{-\delta_0}=0 \, .
\ee
Therefore the expectation value of the operator ${\cal O}_A$ sourced by $A_0$, is given by the expression multiplying its time derivative in the previous equation.

A check on the consistency of these assignments is how they behave when a horizon forms. Thermalization after a global quench in an infinite system only happens at the local level. Namely, for any late but finite time there are sufficiently large regions where non-local observables have not yet achieved thermal values. Such observables, as for example the entanglement entropy, require information from behind the apparent horizon for their holographic determination \cite{AbajoArrastia:2010yt}\cite{Hartman:2013qma}. One-point functions are local observables, which thus should only imply the geometry outside it. We have used constant $t$ slices to translate wall boundary values into QFT couplings. Constant $t$ slices only approach the apparent horizon asymptotically at late times, in the region where it has practically achieved its final value $z_{BH}$. They depart again from it at $z\!>\!z_{BH}$, and finally reach the wall. This implies that indeed, $\langle {\cal O}_0 \rangle$ and $\langle {\cal O}_A \rangle$ do not require information from behind the apparent horizon at any instance of their evolution.

The only non vanishing one-point function associated to a Schwarzchild geometry is that of the stress tensor. Thus other expectation values should tend to zero in the process of gravitational collapse. 
When a horizon emerges, the part of the geometry with $z\!>\!z_{BH}$ gets frozen for observers using the proper time at the AdS boundary. This is implemented by the exponential vanishing of $e^{-\delta}$ in that region. According to the previous assignments both
$\langle {\cal O}_0 \rangle$ and $\langle {\cal O}_A \rangle$ are proportional to $e^{-\delta_0}$, which insures that indeed they tend to zero as a horizon forms. Clearly so does $\langle {\cal O}_\infty \rangle$. 
The evolution of the three observables after a quench generating a horizon, or equivalently leading to thermalization, is shown in Fig.\ref{fig:dev}b.

\vspace{2mm}

We have proposed a simple holographic scenario, easily accessible to numerics, modeling quenches where a relevant coupling changes. A number of checks have been successfully performed. 
We hope that this can help placing holography among the standard tools for studying out of equilibrium physics.

 \vspace{2mm}

{\it  Acknowledgements.} We thank M. Garcia-Perez, R. Emparan and D. Mateos for discussions.
This work was supported by project FPA2015-65480-P and  
Centro de Excelencia Severo Ochoa Programme under grant SEV-2012-0249.

\end{document}